\documentclass[twocolumn,floatfix,showpacs,prd,aps,tightenlines]{revtex4}
\input epsf
\usepackage{graphicx}
\newcommand{\beq}{\begin{equation}}
\newcommand{\beqa}{\begin{eqnarray}}
\newcommand{\eeq}{\end{equation}}
\newcommand{\eeqa}{\end{eqnarray}}

\newcommand{\mnras}{Mon.Not.R.Astr.Soc.}

\begin{document}
\title{
Three-body equations of motion in successive post-Newtonian approximations
} 
\author{Carlos O. Lousto and Hiroyuki Nakano
}
\affiliation{
Center for Computational Relativity and Gravitation, 
School of Mathematical Sciences, \\
Rochester Institute of Technology, 
Rochester, New York 14623, USA
}

\begin{abstract}

There are periodic solutions to the equal-mass three-body 
(and $N$-body) problem 
in Newtonian gravity. The figure-eight solution is one of them. 
In this paper, we discuss its solution in the first and second post-Newtonian
approximations to General Relativity. To do so we derive the
canonical equations of motion in the ADM gauge from
the three-body Hamiltonian. We then integrate those equations
numerically, showing that quantities such as the energy,
linear and angular momenta are conserved down to numerical error.
We also study the scaling of the initial parameters with the 
physical size of the triple system. In this way we can assess
when general relativistic results are important and we determine
that this occur for distances of the order of $100M$, with $M$ the
total mass of the system. For distances much closer than those,
presumably the system would completely collapse due to gravitational
radiation. This sets up
a natural cut-off to Newtonian $N$-body simulations. The method
can also be used to dynamically  provide initial parameters for
subsequent full nonlinear numerical simulations.

\end{abstract}
\maketitle


\section{Introduction}

The closest star to the solar system, Alpha Centauri, is
a triple system, so is Polaris and HD 188753.
Triple stars and black holes are common in globular
clusters~\cite{Gultekin:2003xd, Miller:2002pg}, and
galactic disks. Triple black hole mergers can be formed
in galaxy merger~\cite{Valtonen96} and a triple quasar, 
representing a triple supermassive black hole system has
been recently discovered~\cite{Djorgovski:2007ka}.

Full numerical simulations of black holes made possible
only in the last couple of years have already produced
numerous astrophysically interesting results, among them,
the orbital hangup and respect of the cosmic censorship
hypothesis for spinning black 
holes~\cite{Campanelli:2006uy,Campanelli:2006fg,Campanelli:2006fy},
precession and spin-flips~\cite{Campanelli:2006fy},
and the discovery~\cite{Campanelli:2007ew} of large recoil
velocities in highly-spinning black hole mergers 
up to $4,000$ km/s~\cite{Campanelli:2007cga}.

The 2005 breakthroughs in Numerical 
Relativity~\cite{Pretorius:2005gq,Campanelli:2005dd,Baker:2005vv},
not only provided a solution to the long standing 
two-body problem in General Relativity, but it also 
proved applicable to the black hole - neutron star 
binaries~\cite{Faber:2007dv} and recently 
to the three (and $N$) - black holes systems~\cite{Campanelli:2007ea}.

In general, the solution of three-body problem in Newtonian gravity 
can be chaotic. There are however, periodic orbits in the problem 
of three equal masses on a plane. 
One of the most surprising solution is a figure-eight 
orbit. The three bodies chase each other forever around a fixed
eight-shaped curve. 
This was found first by Moore~\cite{Moore:1993}  
and discussed with the proof of the existence 
in Ref.~\cite{Chenciner:2000}. Heggie~\cite{Heggie:2000}
also estimates the probability for such systems to occur in a galaxy.

Because of effects such as the perihelion shift, 
it was unclear if the figure-eight orbits would exist
in a low post-Newtonian expansion, even if it consist of only
conservative terms. Imai, Chiba and Asada succeeded in obtaining 
the figure-eight solution in a first post-Newtonian order approximation
by finding the general relativistic corrections 
to the Newtonian initial conditions. 
In Ref.~\cite{Chiba:2006ad} they also estimated the 
periodic gravitational waves from this system.

In Ref.~\cite{Imai:2007gn} was used the Euler-Lagrange equations
of motion in an approximation to first post-Newtonian order. In our
paper we instead assume the Hamiltonian formulation to derive the equations
of motion. We start from the Hamiltonian given in Ref.~\cite{Schaefer87}
(with typos corrected in our Appendix). We derive the equations of motion
in this formalism, which are different from those used in 
Ref.~\cite{Imai:2007gn} and have the virtue of explicitly satisfying
the constants of motion of the problem, and thus being more amenable
to numerical integration.

The paper is organized as follows. 
In Section~\ref{sec:EOM}, we summarize the equations of motion 
to be solved numerically in order to obtain the figure-eight orbits. 
The starting point is the three-body Hamiltonian 
in the first post-Newtonian approximation. 
In Section~\ref{sec:1PN}, we discuss 
the initial conditions for the figure-eight solutions.
We study the scaling relation between the orbital radius 
and the linear momenta. From this analysis, we can estimate 
when general relativistic effects are important. 
In Section~\ref{sec:2PN}, we extend our calculation to 
the second post-Newtonian order and
in Section~\ref{sec:DIS}, we summarize the results of this paper 
and discuss some remaining problems. The 2PN three-body Hamiltonian
is explicitly given in the Appendix.
Throughout this paper, we use units in which $c=G=1$. 

\section{Equations of motion}\label{sec:EOM}

As we mentioned in the introduction, the Newtonian configuration
that leads to orbital braid figures can also be obtained
within the Lagrangian approach, in the first post-Newtonian approximation,
by finding the appropriate corrections to the initial data~\cite{Imai:2007gn}.

Here we will consider the Hamiltonian formulation since it generates
equations of motion that conserve the energy, linear and angular momenta.
This is crucial to reach high accuracy in the numerical integrations,
which is needed to keep good track of the orbital motion, that in the
three body problem might be chaotic.

The Hamiltonian ($H=H_N+H_{1PN}+H_{2PN}$) for the three body problem 
in the second post-Newtonian approximation is given next. Note that since
gravitational radiation only enters at $2.5PN$-order and higher, the current
analysis applies to conservative systems.

The Newtonian  Hamiltonian is given by
\begin{eqnarray}
H_N
&=&
\frac{1}{2}\sum_a \frac{p_a^2}{m_a} 
- \frac{1}{2}\sum_{a, b\ne a} \frac{m_a m_b}{r_{ab}} \,,
\label{eq:Hn}
\end{eqnarray}
and to the first post-Newtonian order by
\begin{eqnarray}
H_{1PN}
&=&
-\frac{1}{8}\sum_a m_a \left(\frac{p_a^2}{m_a^2}\right)^2 
\nonumber \\
&&
-
\frac{1}{4}\sum_{a, b\ne a} \frac{m_a m_b}{r_{ab}}
\Big\{ 
  6 \frac{p_a^2}{m_a^2} - 7 \frac{{\bf p}_a \cdot {\bf p}_b}{m_a m_b}
\nonumber \\
&&
  - \frac{\left( {\bf n}_{ab} \cdot {\bf p}_a \right)
  \left( {\bf n}_{ab} \cdot {\bf p}_b \right)}{m_a m_b}
\Big\}
\nonumber \\ &&
+ \frac{1}{2}\sum_{a, b \ne a,c \ne a} \frac{m_a m_b m_c}{r_{ab} r_{ac}} \,,
\label{eq:H1pn}
\end{eqnarray}
where $a,\,b$ and $c$ run over $1,\,2$ and $3$. 
We have used the notations; 
${\bf x}_{ab}={\bf x}_a-{\bf x}_b$, $r_{ab}=|{\bf x}_{ab}|$, 
${\bf n}_{ab}={\bf x}_{ab}/r_{ab}$, $p_a^2={\bf p}_a \cdot {\bf p}_a$ 
and the dot ($\cdot$) means the inner product. 
The Hamiltonian for the second post-Newtonian order is given in the Appendix.

We then obtain the canonical equations
\begin{eqnarray}
({\dot p}_a)_i = - \frac{\partial H}{\partial (q_a)_i} \,, 
\qquad ({\dot q}_a)_i = \frac{\partial H}{\partial (p_a)_i} \,,
\end{eqnarray}
where $i$ denotes $x,\,y$ or $z$.

Explicitly, the equation of motion for the first post-Newtonian order, 
are given for the particle 1 by 
\begin{eqnarray}\label{eq:eom1pn}
{\frac {\partial }{\partial t}}\,{\bf{x}_{1}} 
&=& 
{\displaystyle \frac {{{\bf p}_{1}}}{{m_{1}}}}
- {\displaystyle \frac {1}{2}} \,{\displaystyle \frac {
\,({{\bf p}_{1}}\cdot{{\bf p}_{1}})\,{{\bf p}_{1}}}{{m_{1}}^{3}}} 
\nonumber \\ && 
- {\displaystyle \frac {1}{2}} \,{\displaystyle \frac {{m_{1}}\,{m_{2}}}
{{r_{12}}}} \, 
\left(  \! {\displaystyle \frac {6\,{{\bf p}_{1}}}{{m_{1}}^{2}}}  
- {\displaystyle \frac {7}{2}} \,
{\displaystyle \frac {{{\bf p}_{2}}}{{m_{1}}\,{m_{2}}}} 
\right. \nonumber \\ && \left.  
- {\displaystyle \frac {1}{2}} \,{\displaystyle \frac {
\,({{\bf p}_{2}}\cdot{{\bf x}_{12}} ){{\bf x}_{12}}}
{{m_{1}}\,{m_{2}}
\,{r_{12}}^{2}}}  \!  \right)  
\nonumber \\ && 
+ {\displaystyle \frac {1}{2}} \, \left(  \!  {\displaystyle \frac {7}{2}} 
\,{\displaystyle \frac {{{\bf p}_{3}}}{{r_{31}}}}  
+ {\displaystyle \frac {1}{2}} \,{\displaystyle 
\frac {({{\bf p}_{3}}\cdot{{\bf x}_{31}} )
\,{{\bf x}_{31}}}{{r_{31}}^{3}}}  \!  \right) 
\nonumber \\ && 
- {\displaystyle \frac {1}{2}} \,{\displaystyle \frac {{m_{1}}\,{m_{3}}}
{{r_{31}}}} \, \left(  \! {\displaystyle \frac 
{6\,{{\bf p}_{1}}}{{m_{1}}^{2}}}  
- {\displaystyle \frac {7}{2}} 
\,{\displaystyle \frac {{{\bf p}_{3}}}{{m_{1}}\,{m_{3}}}}  
\right. \nonumber \\ && \left. 
- {\displaystyle \frac {1}{2}} \,{\displaystyle \frac 
{({{\bf p}_{3}}\cdot{{\bf x}_{31}})
\,{{\bf x}_{31}}}{{m_{3}}\,{m_{1}}\,{r_{31}}^{2}}}  \!  \right)  
\nonumber \\ && 
+ {\displaystyle \frac {1}{2}} \, 
\left(  \!  {\displaystyle \frac {7}{2}} 
\,{\displaystyle \frac {{{\bf p}_{2}}}{{r_{12}}}}  
+ {\displaystyle \frac {1}{2}} \,{\displaystyle \frac 
{({{\bf p}_{2}}\cdot{{\bf x}_{12}})\,{{\bf x}_{12}}}
{{r_{12}}^{3}}}  \!  \right) \,,
\\ 
{\frac {\partial }{\partial t}}\,{{\bf p}_{1}} 
&=& - 
\frac{{\bf x}_{12}}{{r_{12}}} \left( {\vrule 
height1.31em width0em depth1.31em} \right. \!  \!  
{\displaystyle \frac {{m_{1}}\,{m_{2}}}
{{r_{12}}^{2}}} 
- {\displaystyle \frac {{m_{2}}\,{m_{1}}\,{m_{3}}}
{{r_{12}}^{2}\,{r_{2, \,3}}}}  
- {\displaystyle \frac {{m_{1}}\,{m_{2}}\,{m_{3}}}{{
r_{12}}^{2}\,{r_{31}}}} 
\nonumber \\ &&  
+ {\displaystyle \frac {1}{2}} {\displaystyle \frac {{m_{1}}\,{m_{2}}}
{{r_{12}}^{2}}} 
 \left( {\vrule height1.31em width0em depth1.31em} \right. \! \!  
{\displaystyle \frac {3\,({{\bf p}_{1}}\cdot{{\bf p}_{1}})}{{m_{1}}^{2}}}  
+ {\displaystyle \frac {3\,({{\bf p}_{2}}\cdot{{\bf p}_{2}})}{{m_{2}}^{2}}}
\nonumber \\ &&  
- {\displaystyle \frac 
{7\,({{\bf p}_{1}}\cdot{{\bf p}_{2}}) }{{m_{1}}\,{m_{2}}}}  
- {\displaystyle \frac 
{({{\bf p}_{1}}\cdot{{\bf x}_{12}})\,({{\bf p}_{2}}\cdot{{\bf x}_{12}})}{{m_{1}}\,{m_{2}}\,{r_{12}}^{2}}}  
\! \! \left. {\vrule height1.31em width0em depth1.31em} \right)  
\nonumber \\ && 
- {\displaystyle \frac {
({{\bf p}_{1}}\cdot{{\bf x}_{12}})\,({{\bf p}_{2}}\cdot{{\bf x}_{12}})}{{r_{12}}^{4}}}  
- {\displaystyle \frac {{m_{2}}\,{m_{1}}^{2}}
{{r_{12}}^{3}}}  
- {\displaystyle \frac {
\,{m_{1}}\,{m_{2}}^{2}}{{r_{12}}^{3}}}  
 \! \! \left. {\vrule 
height1.31em width0em depth1.31em} \right) 
\nonumber \\ && 
+ \frac{{\bf x}_{31}}{{r_{31}}}
 \left( {\vrule 
height1.31em width0em depth1.31em} \right. \!  \!  
{\displaystyle \frac {{m_{3}}\,{m_{1}}}{{r_{31}}^{2}}} 
- {\displaystyle \frac {{m_{3}}\,{m_{1}}\,{m_{2}}}
{{r_{31}}^{2}\,{r_{2, \,3}}}}  
- {\displaystyle \frac {{m_{1}}\,{m_{2}}\,{m_{3}}}
{{r_{12}}\,{r_{31}}^{2}}} 
\nonumber \\ && 
+ {\displaystyle \frac {1}{2}} 
{\displaystyle \frac {{m_{3}}\,{m_{1}}}{{r_{31}}^{2}}} 
 \left( {\vrule height1.31em width0em depth1.31em} 
\right. \!  \! 
{\displaystyle \frac {3\,({{\bf p}_{3}}\cdot{{\bf p}_{3}})}{{m_{3}}^{2}}} 
+ {\displaystyle \frac {3\,({{\bf p}_{1}}\cdot{{\bf p}_{1}})}{{m_{1}}^{2}}} 
\nonumber \\ &&  
- {\displaystyle 
\frac {7\,({{\bf p}_{1}}\cdot{{\bf p}_{3}})}{{m_{1}}\,{m_{3}}}}  
- {\displaystyle \frac {({{\bf p}_{3}}\cdot{{\bf x}_{31}})
\,({{\bf p}_{1}}\cdot{{\bf x}_{31}})}{{m_{3}}\,{m_{1}}
\,{r_{31}}^{2}}}  \! \! \left. {\vrule 
height1.31em width0em depth1.31em} \right) 
\nonumber \\ && 
- {\displaystyle \frac 
{{({\bf p}_{3}}\cdot{{\bf x}_{31}})
\,({{\bf p}_{1}}\cdot{{\bf x}_{31}})}{{r_{31}}^{4}}}  
- {\displaystyle \frac {{m_{3}}\,{m_{1}}^{2}}
{{r_{31}}^{3}}}  
- {\displaystyle \frac {{m_{1}}\,{m_{3}}^{2}}
{{r_{31}}^{3}}}  
 \! \! \left. {\vrule height1.31em width0em depth1.31em} \right) 
\nonumber \\ && 
- {\displaystyle \frac {1}{2}} \,
\left(  \!  
{\displaystyle \frac {({{\bf p}_{2}}\cdot{{\bf x}_{12}})\,{{\bf p}_{1}}}
{{r_{12}}^{3}}}  
+ {\displaystyle \frac {({{\bf p}_{1}}\cdot{{\bf x}_{12}})\,{{\bf p}_{2}}}
{{r_{12}}^{3}}}  \!  \right) 
\nonumber \\ && 
+ {\displaystyle \frac {1}{2}} \,
 \left(  \!  {\displaystyle 
\frac {({{\bf p}_{1}}\cdot{{\bf x}_{31}})\,{{\bf p}_{3}}}
{{r_{31}}^{3}}}  
+ {\displaystyle \frac {({{\bf p}_{3}}\cdot{{\bf x}_{31}})\,{{\bf p}_{1}}}
{{r_{31}}^{3}}}  \!  \right) \,,
\end{eqnarray}
where to obtain the equation of motion for the particle 2 (and 3), 
we change the subscripts as $\{1 \to 2, 2 \to 3, 3 \to 1\}$ 
(and $\{1 \to 3, 2 \to 1, 3 \to 2\}$), respectively.

We solved the above equations numerically for three body problems
using a 10 digits precision implemented in {\it Maple} 10 with typical
runs of a few seconds on a Laptop. 
Since we use the canonical momentum in the calculation, 
the Hamiltonian $H$, the total linear momentum ${\bf P}=\sum {\bf p}_a$ 
and angular momentum ${\bf L}=\sum {\bf x}_a \times {\bf p}_a$ 
are conserved quantities. These represent useful 
checks of the accuracy of the numerical runs.

\section{The first post-Newtonian corrections}\label{sec:1PN}

In the Newtonian case, a figure-eight motion can be obtained 
from the following initial conditions~\cite{Imai:2007gn},
 i.e., the positions ${\bf l}$ and linear momenta ${\bf p}$: 
\begin{eqnarray}
{\bf l} &=& (x_1,\,y_1) = (-x_2,\,-y_2) 
\nonumber \\ &=& (97.00,\,-24.31) \,,
\nonumber \\ 
(x_3,\,y_3) &=& (0,\,0) \,,
\nonumber \\ 
{\bf p}_N &=& (p_3^x,\,p_3^y) 
= (-2p_1^x,\,-2p_1^y) = (-2p_2^x,\,-2p_2^y) 
\nonumber \\ 
&=& (-0.09324,\,-0.08647) \,.
\label{eq:ic}
\end{eqnarray}
Here, we set $m_1=m_2=m_3=m=1$. For the above initial condition, 
the total linear momentum and angular momentum are zero.

At the 1PN order, we also impose the total linear momentum 
${\bf P}=0$ and the total angular momentum ${\bf L}=0$. 
By these conditions, we find that 
each linear momentum is given by the relations
\begin{eqnarray}
{\bf p}_3=-2{\bf p}_1=-2{\bf p}_2 \,.
\end{eqnarray}
Therefore, when we give the positions of the three particles, and 
it is necessary then only to search numerically for ${\bf p}_3$. 
In order to obtain, ${\bf p}_3$, we make some
iterative computations until the figure-eight is reproduced for
a few orbits.

\begin{figure}  \begin{center}
\epsfxsize=8cm
\begin{minipage}{\epsfxsize} 
\epsffile{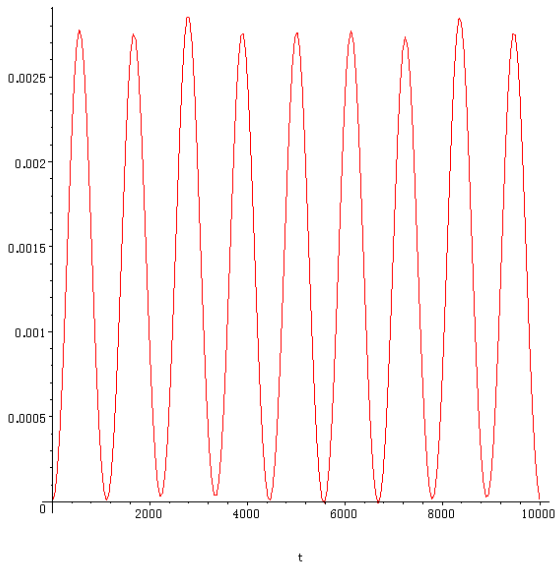} 
\end{minipage}
\caption{The relative error of the Hamiltonian constraint in the Lagrangian
approximation. 
This figure is derived by using the orbit in \cite{Imai:2007gn}.}
\label{fig:H_ICA}
\end{center}
\end{figure}

\begin{figure}  \begin{center}
\vspace{3mm}
\epsfxsize=8cm
\begin{minipage}{\epsfxsize} 
\epsffile{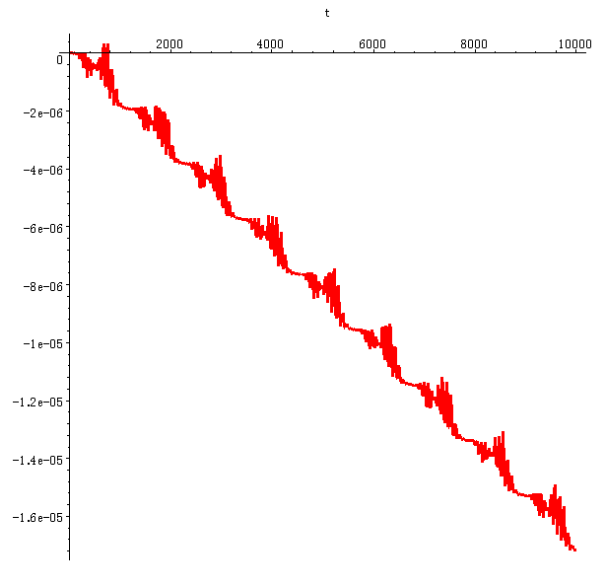} 
\end{minipage}
\caption{The relative error of the Hamiltonian constraint
evaluate by using the orbit derived 
from the Hamiltonian formalism. }
\label{fig:H_H}
\end{center}
\end{figure}

In Figures~\ref{fig:H_ICA} and \ref{fig:H_H}
we show the relative error of the Hamiltonian conservation: 
\begin{eqnarray*}
\Delta H(t) = \frac{H(t)-H(0)}{H(0)} \,. 
\end{eqnarray*}
Figure~\ref{fig:H_ICA} is estimated by 
using the orbit calculated in \cite{Imai:2007gn} and we observe
that they lead to violations of the order of $3\times10^{-3}$. 
While, Figure~\ref{fig:H_H} is derived 
by using the canonical equations derived in our paper and they display
errors of the order of $10^{-6}$, growing linearly in time due to the 
propagation of numerical errors triggered by initial roundoff.

Next we will discuss the scaling behavior of ${\bf p_3}$ 
when we change the initial separation as ${\bf l} \to \lambda{\bf l}$,
and hence the size of the orbit. 
Note that ${\bf p_3} \to \lambda^{-1/2}{\bf p_3}$ 
in the Newtonian limit as can be easily derived from the Hamiltonian 
in Eq.~(\ref{eq:Hn}) or the equations of motion~(\ref{eq:eom1pn}). 

In Table~\ref{tab:results}, we summarize our numerical findings for
the $1PN$ initial conditions for $\lambda$ from 1 to 100. 
We note that ${\bf p}_3$ with $\lambda=1$ is different from the value 
which are derived from the initial velocity of \cite{Imai:2007gn}. 
The value $\theta$ in the table is the inclination angle of the
principal axes. The principal axes of the 1PN figure-eight motion 
are not along the x and y axes~\cite{Imai:2007gn}.

\begin{figure}  \begin{center}
\epsfxsize=8.5cm
\begin{minipage}{\epsfxsize} 
\epsffile{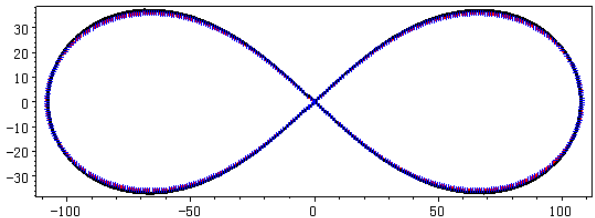} 
\end{minipage}
\caption{Figure-eight motions. We show $\lambda=1$ (solid line)
$\lambda=10$ (dashed line) and $\lambda=100$ (dotted line).}
\label{fig:fig8}
\end{center}
\end{figure}

In Figure~\ref{fig:fig8}, 
the figure-eight rescaled orbits with $\lambda=1,\,10$ and $100$ are shown. 
Here, in order to display the general relativistic effects, 
we have used the coordinates: $(x_a(t)/\lambda,\,y_a(t)/\lambda)$. 
We have chosen here the x-axis as the principal axis. 
We observe that the superposition of the $\lambda=10$ and $\lambda=100$
is suggestive that at those scales the general relativistic effects
are very small 
while for $\lambda <1$ they are dominant, but remainder 
gauge effects may also mask this effect because 
the orbits are not gauge invariant. A cleaner analysis can
be made directly looking at the initial linear momenta scaling.

\begin{table}
\caption{The initial conditions and inclination angle.}
\label{tab:results}
\begin{center}
\begin{tabular}{lccc}
\hline
\hline
$\lambda$ & $(p_3)_x$ & $(p_3)_y$ & $\theta$ \\
\hline
1.00  & -0.09811067089 & -0.09490870640 & 0.01535863098 \\
2.00  & -0.06754964265 & -0.06392246619 & 0.007238984240 \\
5.00  & -0.04209168100 & -0.03934705365 & 0.002786451510 \\
10.00 & -0.02961805051 & -0.02758150399 & 0.001351084509 \\
20.00 & -0.02089989478 & -0.01941808121 & 0.0006871250545 \\
50.00 & -0.01319661317 & -0.01225031026 & 0.0002447024114 \\
100.00 & -0.009328662000 & -0.008654573162 & 0.0001269692928 \\ 
\hline
\end{tabular}
\end{center}
\end{table}

\begin{figure}  \begin{center}
\epsfxsize=8cm
\begin{minipage}{\epsfxsize} 
\epsffile{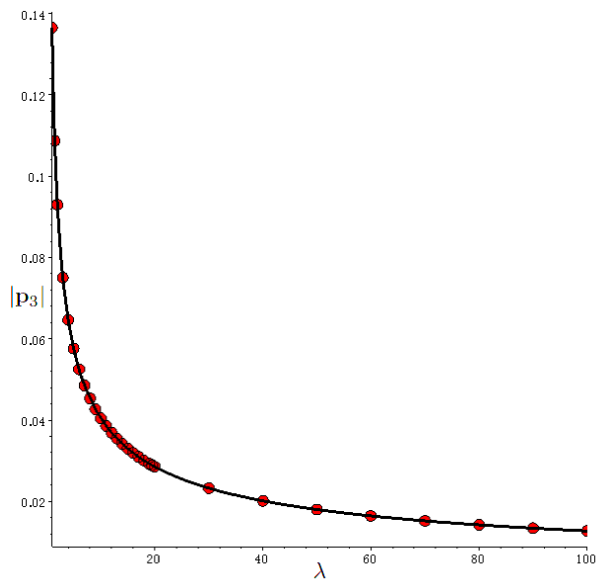} 
\end{minipage}
\caption{$\lambda$-$|{\bf p}_3|$ relation with points obtained numerically.}
\label{fig:LP}
\end{center}
\end{figure}

\begin{figure}  \begin{center}
\vspace{5mm}
\epsfxsize=8cm
\begin{minipage}{\epsfxsize} 
\epsffile{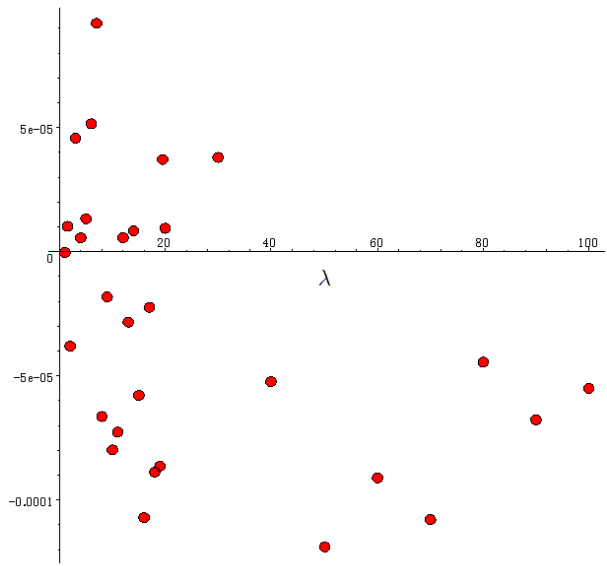} 
\end{minipage}
\caption{The relative error of the fitting.}
\label{fig:LPe}
\end{center}
\end{figure}

By using the results of the runs in Table~\ref{tab:results},
we propose a fitting formula for $|{\bf p}_3|$ inspired again
in the 1PN Hamiltonian or the equations of motion 
\begin{widetext}
\begin{eqnarray}
|{\bf p}_3|_{\rm fit}(\lambda) = \sqrt{
\frac{0.01617387234}{\lambda}+\frac{0.002042558971}{\lambda^2}
+\frac{0.0004169461512}{\lambda^3} } \,.
\label{eq:1PNfit}
\end{eqnarray}
\end{widetext}
In Figure~\ref{fig:LP}, we show the fitting function and in
Figure~\ref{fig:LPe} we display the
relative error 
$|{\bf p}_3|-|{\bf p}_3|_{\rm fit}/|{\bf p}_3|$ \,,
consistent with the form of an error generated in the numerical calculation.

Independently in the Newtonian calculations, 
the $\lambda$-$|{\bf p}_3|$ relation can be obtained 
from the initial condition in Eqs.~(\ref{eq:ic}) as 
\begin{eqnarray}
|{\bf p}_3|_{\rm N}(\lambda) = \frac{0.1271642973}{\lambda^{1/2}} \,.
\end{eqnarray}
Note that relative difference $|{\bf p}_3|$ 
between the Newtonian and the first post-Newtonian calculations:
\begin{eqnarray*}
\frac{|{\bf p}_3|_{\rm fit}(\lambda)-|{\bf p}_3|_{\rm N}(\lambda)}
{|{\bf p}_3|_{\rm N}(\lambda)} \,,
\end{eqnarray*}
is 7\% for $\lambda=1$, 0.6\% for $\lambda=10$ 
and 0.07\% for $\lambda=100$.

\section{Second post-Newtonian corrections}\label{sec:2PN}

\begin{figure}  \begin{center}
\epsfxsize=8cm
\begin{minipage}{\epsfxsize} 
\epsffile{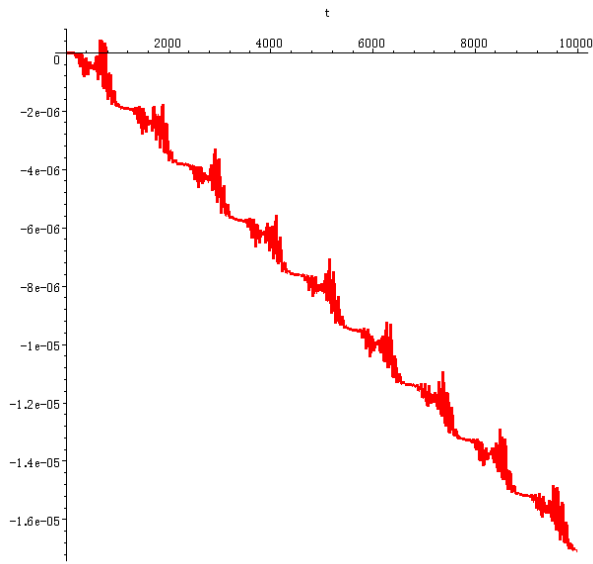} 
\end{minipage}
\caption{The relative error of the Hamiltonian conservation 
for the second post-Newtonian order calculations.}
\label{fig:H_H2PN}
\end{center}
\end{figure}

It is interesting to verify if this kind of orbits also exists in
the second post-Newtonian approximation to General Relativity, since
they incorporate further effects of the curvature, but yet not
gravitational radiation.
The calculations are done by using the same method as for the first 
post-Newtonian order. 
In Table~\ref{tab:results2PN}, we summarize the initial conditions 
for each $\lambda$ from 1 to 100. 
We show the numerical errors as measured through the Hamiltonian
non-conservation in Figure~\ref{fig:H_H2PN}.

We find that we can approximate $|{\bf p}_3|$
by the fitting formula
\begin{widetext}
\begin{eqnarray}
|{\bf p}_3|_{\rm fit}(\lambda) = \sqrt{
\frac{0.01617654493}{\lambda}
+\frac{0.002017242451}{\lambda^2}
+\frac{0.0002017242451}{\lambda^3} 
+\frac{0.0001054698539}{\lambda^4} }
\,.
\label{eq:2PNfit}
\end{eqnarray}
\end{widetext}
There is a significant difference between the coefficient of $1/\lambda^3$ 
in Eqs.~(\ref{eq:1PNfit}) and (\ref{eq:2PNfit}). This is due
to second post-Newtonian corrections entering in this coefficient,
as we can verify from the form of the Hamiltonian.

\begin{figure}  \begin{center}
\epsfxsize=8cm
\begin{minipage}{\epsfxsize} 
\epsffile{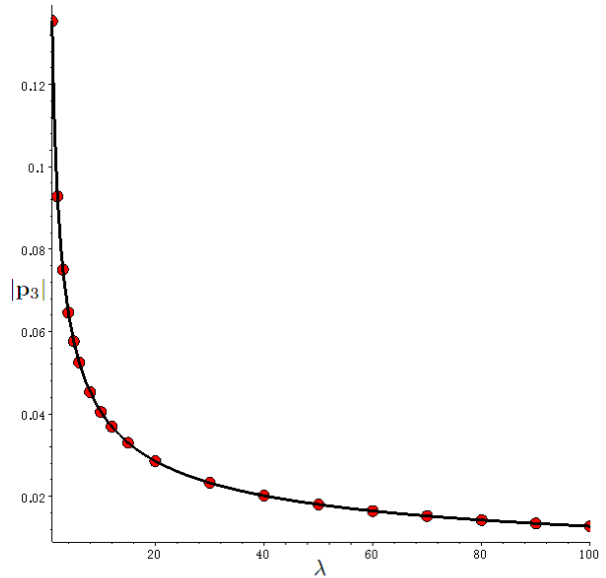} 
\end{minipage}
\caption{$\lambda$-$|{\bf p}_3|$ relation for the second post-Newtonian case. 
The points are obtained numerically.}
\label{fig:LP2}
\end{center}
\end{figure}

\begin{figure}  \begin{center}
\epsfxsize=8cm
\begin{minipage}{\epsfxsize} 
\epsffile{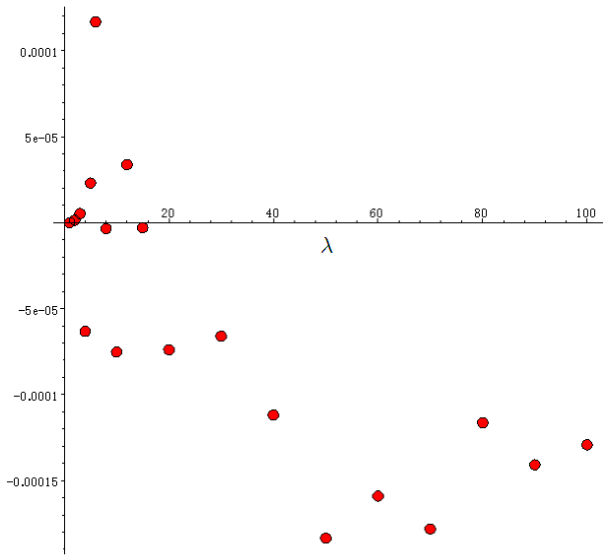} 
\end{minipage}
\caption{The relative fitting error for the second post-Newtonian case.}
\label{fig:LPe2}
\end{center}
\end{figure}

In Figure~\ref{fig:LP2}, we show the fitting function while its 
relative error is given in Figure~\ref{fig:LPe2}.

\begin{table}
\caption{The initial conditions and inclination angle for the second post-Newtonian case.}
\label{tab:results2PN}
\begin{center}
\begin{tabular}{lccc}
\hline
\hline
$\lambda$ & $(p_3)_x$ & $(p_3)_y$ & $\theta$ \\
\hline
1.00  & -0.09759146109 & -0.09386471063 & 0.01335212441 \\
2.00  & -0.06746813797 & -0.06375625776 & 0.006775950067 \\
5.00  & -0.04208326266 & -0.03933131483 & 0.002713363325 \\
10.00 & -0.02961805051 & -0.02757874584 & 0.001340868765 \\
20.00 & -0.02089780479 & -0.01941808121 & 0.0006733410290 \\
50.00 & -0.01319661317 & -0.01225031026 & 0.0002447024114 \\
100.00 & -0.009328662000 & -0.008654573162 & 0.0001269692928 \\ 
\hline
\end{tabular}
\end{center}
\end{table}

\begin{figure}  \begin{center}
\epsfxsize=8cm
\begin{minipage}{\epsfxsize} 
\epsffile{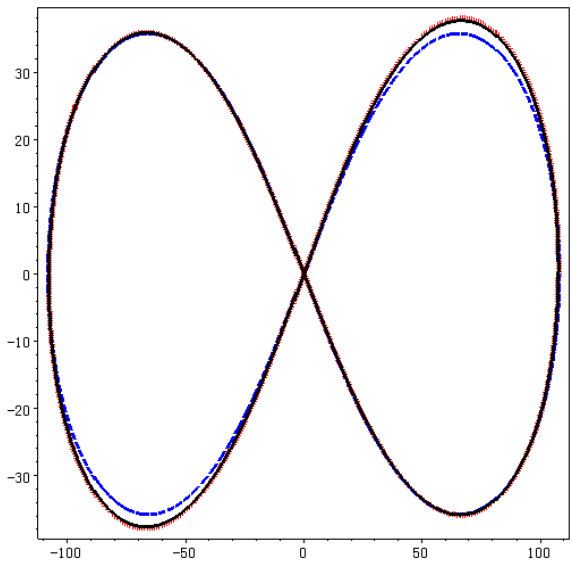} 
\end{minipage}
\caption{Comparison of figure-eight motions for $\lambda=1$. The solid, 
dotted and dashed lines show the 2PN, 1PN and Newtonian results, 
respectively.}
\label{fig:N12}
\end{center}
\end{figure}

Finally, we summarize the results by showing the difference between
the Newtonian, first and second post-Newtonian results 
in Figure~\ref{fig:N12}. The second post-Newtonian effect is small but
clearly not negligible for $\lambda=1$. 

\section{Discussion}\label{sec:DIS}

In this paper we have used the figure-eight orbits as a theoretical
lab to test the properties of the low post-Newtonian expansions of
General Relativity. We have found that those closed orbits exists
for three (and presumably $N$) bodies. We have provided an
improved first-post-Newtonian order formalism for deriving the
equations of motion that satisfy the Hamiltonian (the linear and 
angular momenta) constraint to round-off
error. The subsequent numerical evolution is well behaved during 
for more than $t\sim10,000m$. 
We have also extended this analysis to the $2PN$ corrections,
still giving a conservative system of equations. In the process
of finding the figure-eight solutions by trial of different
initial momenta we also showed (numerically) the stability of the orbit against
small perturbations.

This method is particularly useful to determine, dynamically
(as an alternative to determine them through families of initial 
data~\cite{Campanelli:2005kr}), initial orbital parameters for
subsequent full numerical evolution~\cite{Campanelli:2007ea},
when the holes are close enough that general relativistic effects can
no longer be ignored. Note that our method fully takes into account
the three-body post-Newtonian interactions unlike other simulations
that approximate the problem in successive two-body 
problems~\cite{Aarseth:2007wv}.

It is interesting to note here that the scaling fits~(\ref{eq:2PNfit})
give a practical way to determine when relativistic or Newtonian
approaches are appropriate. For $\lambda=1$ we have that the ratio
of the first coefficient, $0.01617654493$ (Newtonian) to the second coefficient
$0.002017242451$ first-post-Newtonian is nearly $0.12/\lambda$ 
and the second coefficient to
the third one $0.0002463605227$ (dominated by second-post-Newtonian) 
is also approximately
$0.12/\lambda$. This indicates that post-Newtonian corrections are important.
For $\lambda=1$ the distance between the initial bodies is $200m$, 
what indicates that for nearly $67M$ with $M\approx3m$ the total mass of the
system has strong post-Newtonian effects. For $\lambda \gg 1$ 
Newtonian gravity should describe the system accurately, while
for $\lambda<1$ general
relativistic effects should be very important, eventually leading to
the total collapse of the system. It is interesting to remark here
that most of the $N$-body codes use some sort of regularization of
the Newtonian gravity for very close encounters \cite{aarseth-03}, instead the natural
way to regularize these close encounters \cite{Campanelli:2007ea} is given by the General Theory
of Relativity, and as we show here, the post-Newtonian corrections are already
non-negligible at separations of the order of $100M$. In any case, for most
of the astrophysical encounters this is way too short distance, but it 
can obviously be reached in systems involving black holes and neutron stars.

\acknowledgments

We would like to thank H.-P.~Bischof, M.~Campanelli, A.~Gualandris, 
D.~Merritt, D.~Ross and Y.~Zlochower for useful discussions. 
This is supported by JSPS for Research Abroad (HN)
and by the NSF through  grants PHY-0722315,  PHY-0701566, PHY-0714388,
and PHY-0722703.


\appendix 

\section{the second post-Newtonian three-body Hamiltonian}\label{app:2pn}

In this appendix, we give explicitly the Hamiltonian for the 
three body problem at second post-Newtonian order in the ADM gauge 
since there are some typos in the summation of~\cite{Schaefer87}. 
The equations of motion used in our paper can be derived straightforwardly 
from this Hamiltonian, but are too cumbersome to write down here.

\begin{widetext}

\begin{eqnarray}
H_{2PN}
&=&
\frac{1}{16} \sum_{a} m_a \left(\frac{p_a^2}{m_a^2}\right)^3 
+
\frac{1}{16} \sum_{a, b\ne a}  
\frac{m_a m_b}{r_{ab}}
\Big\{
  10 \left(\frac{p_a^2}{m_a^2} \right)^2
  - 11 \frac{p_a^2 p_b^2}{m_a^2 m_b^2}
  - 2 \frac{\left( {\bf p}_a \cdot {\bf p}_a \right)^2}{m_a^2 m_b^2}
\nonumber \\
&&
  + 10 \frac{p_a^2 \left( {\bf n}_{ab} \cdot {\bf p}_b \right)^2}{m_a^2 m_b^2}
  - 12 \frac{\left( {\bf p}_a \cdot {\bf p}_b \right)
    \left( {\bf n}_{ab} \cdot {\bf p}_a \right)
    \left( {\bf n}_{ab} \cdot {\bf p}_b \right)}{m_a^2 m_b^2}
  - 3 \frac{ \left( {\bf n}_{ab} \cdot {\bf p}_a \right)^2
    \left( {\bf n}_{ab} \cdot {\bf p}_b \right)^2}{m_a^2 m_b^2}
\Big\}
\nonumber \\ && 
+
\frac{1}{8} \sum_{a, b\ne a, c\ne a}  
\frac{m_a m_b m_c}{r_{ab}\, r_{ac}}
\Big\{
  18 \frac{p_a^2}{m_a^2} 
  +14 \frac{p_b^2}{m_b^2} 
  -2 \frac{\left( {\bf n}_{ab}\cdot {\bf p}_{b} \right)^2}{m_b^2}
  -50 \frac{{\bf p}_{a}\cdot {\bf p}_{b}}{m_a m_b}
  +17 \frac{{\bf p}_{b}\cdot {\bf p}_{c}}{m_b m_c}
\nonumber \\
&& 
  - 14 \frac{\left( {\bf n}_{ab}\cdot {\bf p}_{a} \right)
    \left( {\bf n}_{ab}\cdot {\bf p}_{b} \right)}{m_a m_b}
  + 14 \frac{\left( {\bf n}_{ab}\cdot {\bf p}_{b} \right)
    \left( {\bf n}_{ab}\cdot {\bf p}_{c} \right)}{m_b m_c}
  + {\bf n}_{ab}\cdot {\bf n}_{ac} 
    \frac{\left( {\bf n}_{ab}\cdot {\bf p}_{b} \right)
    \left( {\bf n}_{ac}\cdot {\bf p}_{c} \right)}{m_b m_c}
\Big\}
\nonumber \\
&& 
+
\frac{1}{8} \sum_{{a, b\ne a, c\ne a}}  
\frac{m_a m_b m_c}{r_{ab}^2}
\Big\{
  2 \frac{\left( {\bf n}_{ab}\cdot {\bf p}_{a} \right)
    \left( {\bf n}_{ac}\cdot {\bf p}_{c} \right)}{m_a m_c}
  + 2 \frac{\left( {\bf n}_{ab}\cdot {\bf p}_{b} \right)
    \left( {\bf n}_{ac}\cdot {\bf p}_{c} \right)}{m_a m_c}
\nonumber \\
&& 
  + 5 {\bf n}_{ab}\cdot {\bf n}_{ac} \frac{p_c^2}{m_c^2}
  - {\bf n}_{ab}\cdot {\bf n}_{ac} 
    \frac{ \left( {\bf n}_{ac}\cdot {\bf p}_{c} \right)^2 }{m_c^2}
  - 14 \frac{\left( {\bf n}_{ab}\cdot {\bf p}_{c} \right)
    \left( {\bf n}_{ac}\cdot {\bf p}_{c} \right)}{m_c^2}
\Big\}
\nonumber \\
&& 
+ 
\frac{1}{4} \sum_{a, b\ne a}
\frac{m_a^2 m_b}{r_{ab}^2}
\Big\{
  \frac{p_a^2}{m_a^2} + \frac{p_b^2}{m_b^2} 
  - 2 \frac{ {\bf p}_a \cdot {\bf p}_b }{m_a m_b}
\Big\}
\nonumber \\
&& 
+
\frac{1}{2} \sum_{a, b\ne a, c\ne a,b}
\frac{m_a m_b m_c}{\left(r_{ab} + r_{bc} + r_{ca}\right)^2}
  ( n^i_{ab} + n^i_{ac} ) ( n^j_{ab} + n^j_{cb} ) 
\Big\{  
  8 \frac{p_{ai}p_{cj}}{m_a m_c}
  -16 \frac{p_{aj}p_{ci}}{m_a m_c}
\nonumber \\
&& 
  +3 \frac{p_{ai}p_{bj}}{m_a m_b}
  +4 \frac{p_{ci}p_{cj}}{m_c^2}
  + \frac{p_{ai}p_{aj}}{m_a^2}
\Big\}
\nonumber \\
&& 
+
\frac{1}{2} \sum_{a, b\ne a, c\ne a,b}
\frac{m_a m_b m_c}{\left(r_{ab} + r_{bc} + r_{ca}\right) r_{ab}}
\Big\{
  8 \frac{{\bf p}_{a} \cdot {\bf p}_{c} 
    -\left( {\bf n}_{ab} \cdot {\bf p}_a \right)
    \left( {\bf n}_{ab} \cdot {\bf p}_c \right)}{m_a m_c}
\nonumber \\
&& 
  -3 \frac{{\bf p}_{a} \cdot {\bf p}_{b}
    -\left( {\bf n}_{ab} \cdot {\bf p}_a \right)
    \left( {\bf n}_{ab} \cdot {\bf p}_b \right)}{m_a m_b}
  -4 \frac{p_c^2 - \left( {\bf n}_{ab} \cdot {\bf p}_c \right)^2}{m_c^2}
  -\frac{p_a^2 - \left( {\bf n}_{ab} \cdot {\bf p}_a \right)^2}{m_a^2}
\Big\}
\nonumber \\
&& 
-\frac{1}{2} \sum_{a, b\ne a, c \ne b}
\frac{m_a^2 m_b m_c }{r_{ab}^2\, r_{bc}}
-\frac{1}{4} \sum_{a, b\ne a, c \ne a}
\frac{m_a m_b m_c^2 }{r_{ab}\, r_{ac}^2}
+\frac{1}{2} \sum_{a, b\ne a}
\frac{m_a^3 m_b }{r_{ab}^3}
\nonumber \\ 
&& 
-\frac{3}{4} \sum_{a, b\ne a, c\ne a}
\frac{m_a^2 m_b m_c}{r_{ab}^2\, r_{ac}}
-\frac{3}{8} \sum_{a, b\ne a, c\ne a,b}
\frac{m_a^2 m_b m_c}{r_{ab}\, r_{ac}\, r_{bc}}
+\frac{3}{8} \sum_{a, b\ne a}
\frac{m_a^2 m_b^2}{r_{ab}^3}
\nonumber \\ 
&& 
-\frac{1}{64}\sum_{a, b\ne a, c\ne a,b}
\frac{m_a^2 m_b m_c}{r_{ab}^3\, r_{ac}^3\,r_{bc}}
\big\{
18r_{ab}^2r_{ac}^2-60r_{ab}^2r_{bc}^2-24r_{ab}^2r_{ac}(r_{ab}+r_{bc})
+60r_{ab}r_{ac}r_{bc}^2+56r_{ab}^3r_{bc}
\nonumber \\ &&
-72r_{ab}r_{bc}^3+35r_{bc}^4+6r_{ab}^4
\big\}
-
\frac{1}{4} \sum_{a, b\ne a}  
\frac{m_a^2 m_b^2}{r_{ab}^3} \,.
\end{eqnarray}

\end{widetext}


\end{document}